# Coherent room-temperature dipole synchronization in nanocavity sheets

Rakesh Arul[1*], Piper Fowler-Wright[2,3], Lille Borresen[1], Brendon W. Lovett[2], Jonathan Keeling[2*], Jeremy J. Baumberg[1*]

[1] Nanophotonics Centre, Dept. of Physics, Cavendish Laboratory, University of Cambridge, Cambridge, CB3 0HE, UK
[2] SUPA, School of Physics and Astronomy, University of St. Andrews, St. Andrews, KY16 9SS, United Kingdom
[3] Department of Chemistry and Biochemistry, University of California San Diego, La Jolla, CA 92093, USA
* e-mail: ra554@cam.ac.uk, jmjk@st-andrews.ac.uk, jjb12@cam.ac.uk

**Plasmonic nanocavities enable the synchronization of spatially distant emissive dipoles through strong near-field coupling in sub-nm gaps. We report formation of a room-temperature synchronized dipole state in locally-ordered plasmonic nanogap 2D arrays under non-resonant continuous-wave pumping. Unlike lasers, photonic Bose-Einstein condensates, or exciton-polariton condensates, this system exhibits spatial coherence across the dipoles, while rapid radiative and non-radiative emission suppresses temporal photon coherence. A change of behaviour is observed with increasing pumping, marked by the spatial spread of $g^{(1)}$ coherence, but without spectral narrowing or directional emission. This driven-dissipative system exhibits fast temporal coherence decay and complex spatial correlations, offering a new platform for studying synchronization at room temperature. Combining ultralow mode volumes, high Purcell enhancement, and scalable ambient operation, it opens pathways for novel photonic and quantum technologies.**

Synchronization is ubiquitous, from starling flocks to macroscopic quantum states, and plays a pivotal role in many phenomena including Huygens clocks, exciton-polariton condensation, lasing, superradiance, coherence in biological and chemical systems, and for quantum resources[1-9]. In optical dipole arrays, oscillating charges interact through electromagnetic fields. This interaction strengthens when emitters are coupled via an optical cavity. The 'good-cavity' limit occurs when the photon residence time exceeds the dipole lifetime. In this regime, stimulated emission dominates to produce phase-locked lasing and sharp spectral output. Conversely, the 'bad-cavity' limit couples many narrow-linewidth emitters through a lossy cavity. This regime shares a kinship with Huygens clocks but remains largely unexplored[10,11].

Traditional microcavities typically operate in the good-cavity limit. These systems support exciton-polariton[12,13], photon[14] and plasmonic[15,16] Bose-Einstein condensation[17], as well as parametric oscillation[18]. Such microcavity polaritons enable applications in switching[26], optical computing[27], study of nonequilibrium physics[28,29], and control of chemistry[30]. However, most current designs require microfabrication and cryogenic cooling. They lack the advantages of scalable self-assembly and ambient operation. Plasmonic condensates exist[19,20], but fail to leverage plasmonics' key advantage: extreme optical field enhancements from sub-nm gap confinement[21]. Furthermore, these systems often suffer from low emitter gain and material degradation[22-24] under pulsed excitation[12,25].

Here, we demonstrate a truly room-temperature continuously-driven synchronized dipole state in a self-assembled 2D plasmonic nanogap array. Organic dye emitter molecules embedded within coupled sub-nm gaps form a system of emissive dipoles interacting strongly with a collective optical mode under continuous incoherent driving and strong dissipation. This leads to nonlinear synchronization between emitters, and to emergent extended spatial coherence with short temporal coherence. This driven-dissipative system differs from conventional and random lasers, Bose-Einstein condensates, and superradiant phases, offering a platform to explore how rapidly synchronization develops between emitters. Remarkably, here synchronization arises spontaneously despite high dissipation, incoherent pumping, and room-temperature operation. Tuneable dipole densities and interaction strengths open pathways to study driven-dissipative dynamics in new regimes, where large cavity-dipole coupling rates and loss rates (≫kT) enable collective behaviour at room temperature.

We assemble methylene blue (MB) molecules inside 0.9 nm nanogaps of a close-packed 2D array[31] of 80 nm gold nanoparticles (NPs) (Fig. 1a-c). Cucurbit[7]uril (CB[7]) molecular scaffolds force MB dipoles to be perpendicular to the metal facets, aligning them with the optical field (Fig. 1b, blue arrow). This allows out-of-plane direct excitation of MB emitters by a focused laser, as well as imaging of their emission (after filtering out all pump laser light).

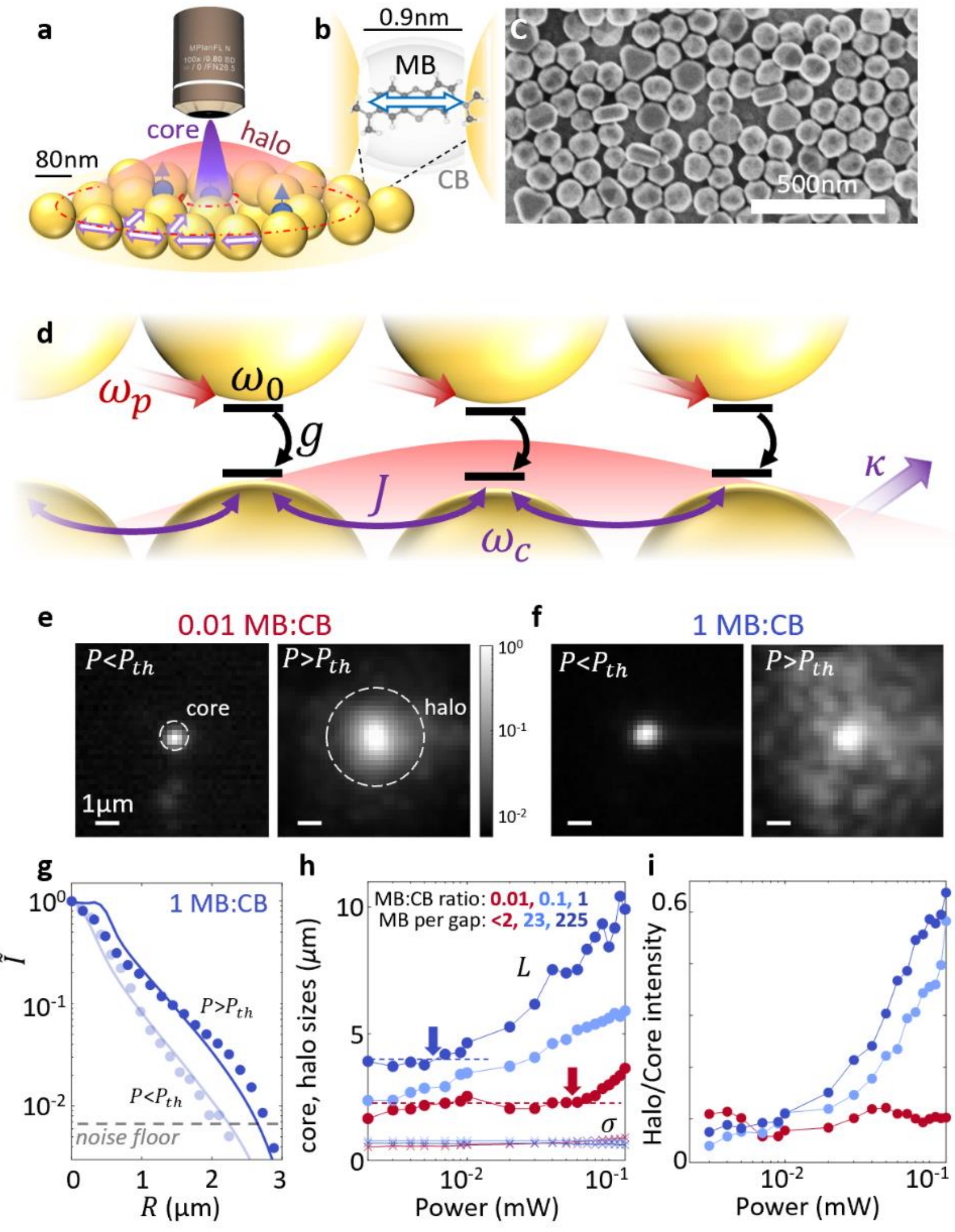

**Fig.1. Synchronization of dipoles in plasmonic nanogaps.** (**a**) Emission from array of nanoparticles pumped by tightly-focused 633 nm laser exciting core region which feeds surrounding halo. (**b**) Methylene blue (MB) emitter within CB[7] host forms optical dipole (blue) in nanogap between gold nanoparticles. (**c**) SEM of array of NPs. (**d**) Each emitter is incoherently pumped by a laser of frequency $\omega_p$ (red arrows) and after non-radiative relaxation emits into the cavity mode with light-matter coupling strength $g$. The cavity is composed of the collective plasmonic nanocavities of individual

frequency $\omega_c$ coupled with plasmon-plasmon coupling strength $J$. The cavity leaks light into the environment at loss rate $\kappa$, which is imaged. (**e,f**) Imaged normalized emission intensity maps below ($P$=2µW) and above ($P$=130µW) the threshold pump power $P_{th}$ (indicated by arrows in Fig. 1h, at which the nonlinear rise begins), for (e) 0.01 and (f) 1 MB:CB ratios. Core and halo regions indicated with dotted circles. All images share the same scale bar. (**g**) Radially averaged normalized intensity for 1 MB:CB below ($P$=2µW) and above ($P$=130µW) threshold. Solid lines are fits to theoretical model (see text). (**h**) Sizes of core ($\sigma$ = full-width-at-half-maximum) and halo ($L$ = exponential decay length). (**i**) Ratio of total emitted intensity of the halo region to the core region *vs* pump power for different MB:CB ratios.

The nanogaps support highly confined optical modes determined by the NP diameter, gap size, and refractive index. Coupling identical nanogaps forms delocalized plasmonic modes across >10 NPs (SI Section S2), offering unusual properties: high light-matter coupling strength $g$ (collectively >100meV), high photon loss $\kappa$ (>100meV), and strong radiative leakage. Self-assembly with MB and CB[7] controls the dipole density per gap $\Xi$ (via dye loading) and self-aligns all dipoles along the local optical near-field, overcoming typical misalignment of randomly-oriented dipoles with microcavity optical fields. Furthermore, dipole-dipole coupling between adjacent nanogaps withstands disorder in the chains[32] (SI Fig. S2).

This system behaves as collective dipoles that couple to each gap plasmonic mode with strength $g$, and plasmonic modes mutually-coupling with hopping strength $J$ in a tight-binding model (Fig. 1d). This model accounts for most plasmonic interaction, differing only slightly from the exact plasmonic chain dispersion (SI Fig. S5). The coupled plasmonic modes exhibit low Q due to fast radiative ($\kappa$) and non-radiative ($\Gamma$) decay, providing a platform for observing dipole synchronization and collective dynamics.

**Spatial Halo Formation**

The samples are pumped at high energy (633nm), giving Stokes-shifted emission around $\lambda$=700nm after incoherent vibrational relaxation. At low pump powers the emission spatially matches the Gaussian pump profile. However as pump powers increase, a dramatic expansion of the emissive region occurs, extending well beyond the 0.5µm diameter pump (Fig.1e,f), forming a 'halo' whose intensity grows with $\Xi$ (Fig.1f). Radially-averaged images (Fig.1g) decompose into a fixed Gaussian core (Fig.1h), and an exponentially decaying larger radial halo. With increasing pump power, the emission from halo dipoles dominates over the core dipoles (Fig. 1i), which is unusual and not expected from ordinary emission where the Gaussian core should remain fixed in size.

To understand the emergence of non-local correlations in the plasmonic system we first model a 1D tight-binding[33] chain of $M$ sites, each site (position $r_n$) hosting $x = \{1, \ldots, \Xi\}$ two-level dipoles of frequency $\omega_0$ (Pauli matrices $\sigma_{nx}^{\alpha}$) interacting with $M$ plasmonic modes $\omega_c$ i.e. photons (bosonic operator $a_n^{\dagger}$ delocalised over the chain) with detuning $\Delta = \omega_0 - \omega_c$. The emitters are subject to incoherent pumping $\Gamma_n^{\uparrow}$, decay $\Gamma^{\downarrow}$ and dephasing $\Gamma^{z}$, as well as pair annihilation (within each gap) at rate $\Gamma^{ee}$. The excitation rate $\Gamma_n^{\uparrow}$ imprints the Gaussian pump profile at the centre of the chain. This corresponds to a master equation defined by:

$$H = \sum_{n,x} \frac{\Delta}{2} \sigma_{nx}^{z} - J \sum_{n} \left( a_{n+1}^{\dagger} a_n + H.c. \right) + g \sum_{n,x} \left( a_n \sigma_{nx}^{+} e^{-ikr_n} + H.c. \right) \tag{1}$$

$$\partial_t \rho = -i[H, \rho] + \sum_{n,x} \left\{ \kappa L[a_n] + \Gamma_n^{\uparrow} L[\sigma_{nx}^{+}] + \Gamma^{\downarrow} L[\sigma_{nx}^{-}] + \Gamma^{z} L[\sigma_{nx}^{z}] \right\} + \Gamma^{ee} \sum_{n,x,y} L\left[\sigma_{nx}^{-} \sigma_{ny}^{-}\right] \tag{2}$$

where $L[X] = X\rho X^{\dagger} - \{X^{\dagger}X, \rho\}/2$. To describe this many-body system, we use a second-order cumulant expansion[34,35] for developing light-matter $P_{nm} = \langle a_n \sigma_m^{+} \rangle$ and emitter-emitter $C_{nm,xy} = \langle \sigma_{nx}^{+} \sigma_{my}^{-} \rangle$ correlations, in addition to equations for photon-photon correlations $N_{nm} = \langle a_n^{\dagger} a_m \rangle$, emitter inversion $S_n = \langle \sigma_n^{z} \rangle$, and local emitter spin correlations $Z_n = \langle \sigma_{nx}^{z} \sigma_{ny}^{z} \rangle$. Further details of the tight-binding model and second-order cumulant expansion are provided in SI Section S3. This approach allows for efficient simulation of the multimode many-molecule system by recognising emitters in a single gap have the same on-site average properties ($S_n$) whilst crucially including the coherences that exist not only across gaps but also between different emitters in the same gap.

Solving for steady-state photon populations $N_{nn}$ and total emission $I \propto N_{tot} = \sum_n N_{nn}$ under increasing pump strength $\Gamma_n^{\uparrow}$ shows linear, nonlinear, and halo-dominated regimes. Increasing the pump amplitude $\Gamma_{n=0}^{\uparrow}$ reproduces the observed spatial photon emission profile $N_{nn}$ (Fig. 1g). This predicted spatial profile (solid lines) shows the expansion of emission when transitioning between regimes. Disorder in our system gives slightly different realisations of the brighter spots (Fig.1e) localised on the scale of our resolution, highlighting the role of local optical transport between nanogaps, out-scattering from these gaps, and the dependence on local NP gap arrangement (Fig.1c). Despite this, robust synchronization is still observed even in the presence of disorder and dissipation[7,36,37].

**Spatio-temporal coherence**

Further evidence of unusual, synchronized emission is provided by the coherence recorded from a Michelson interferometer (Fig.2a). By inverting the emission image in one arm of the interferometer, interference fringes are recorded from increasingly laterally-separated ($R$) emissive regions[38,39] (Fig.2b). This enables the spatial coherence $g^{(1)}(R)$ to be extracted from the visibility of interference fringes (SI Section S4) as a function of the relative time delay ($\tau$).

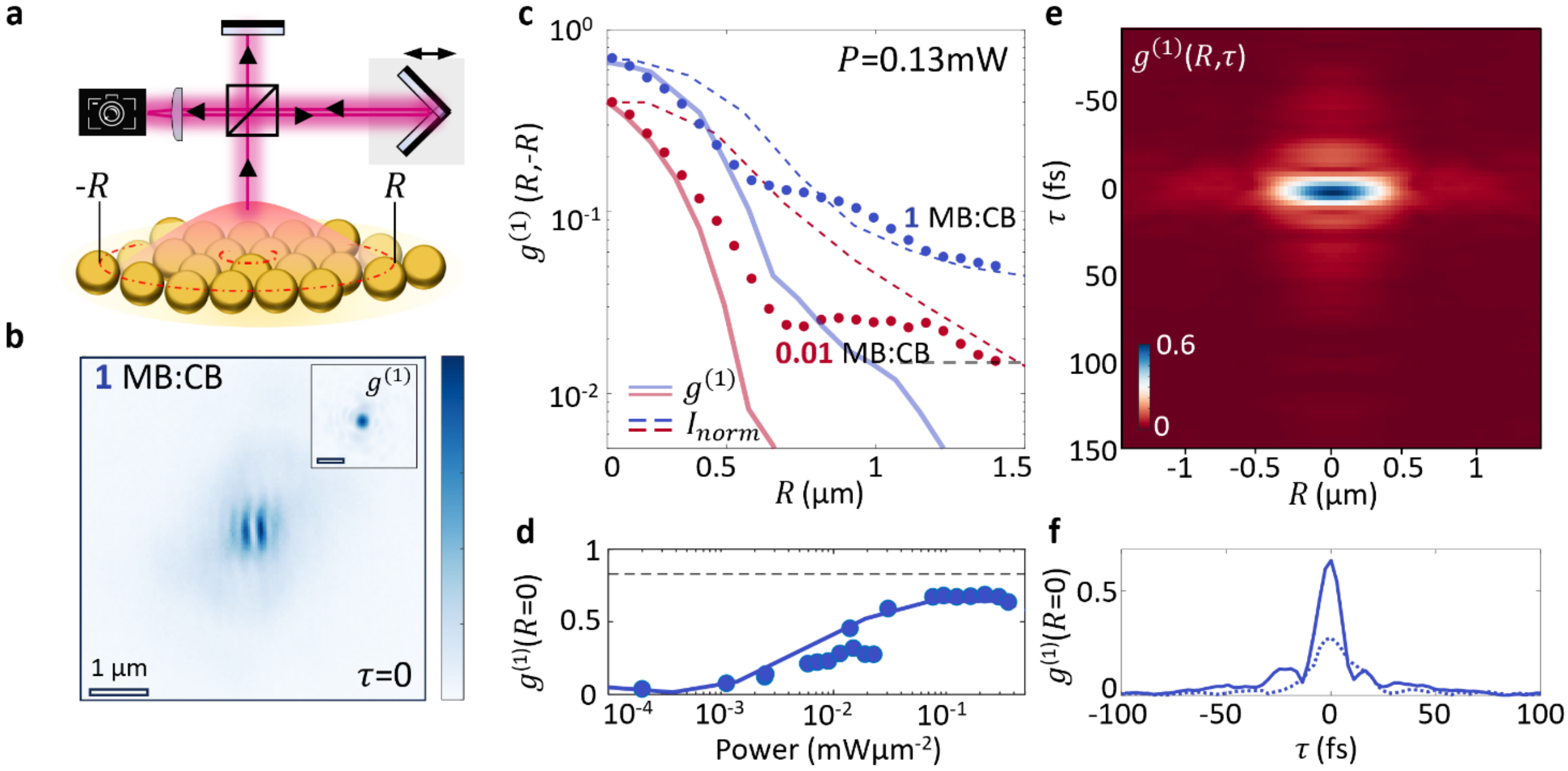

**Fig.2. Spatial coherence of light emission.** (**a**) Flipped Michelson interferometer measuring spatial visibility fringes of emission from the 2D NP layer. (**b**) Emission fringes from MB:CB = 1 sample at zero-delay ($\tau$=0). Inset shows extracted image of $g^{(1)}$ spatial coherence, scale bar as in main image. (**c**) Decay of $g^{(1)}$ spatial coherence over radially averaged

distance from pump spot (points) for MB:CB = 0.01 and 1, together with intensity profiles (dotted lines) and simple theoretical prediction (solid lines). (**d**) Coherence $g^{(1)}$ at core centre *vs* pump power for MB:CB = 1, showing smooth crossover. Solid line is the theoretical model calculation (see SI section S3), dotted line indicates measured coherence of pump laser. (**e**) Spatio-temporal $g^{(1)}(R, \tau)$ decay for MB:CB = 1, with (**f**) extracted time decay at the centre of the pumped region $g^{(1)}(R{=}0,\tau)$ for pump power $P$=0.2 mW (solid line) and 0.02 mW (dotted line).

At zero delay, $g^{(1)}(R,\tau{=}0)$ initially spans half the pump width but extends significantly into the halo region (Fig. 2c). As pump power increases, $g^{(1)}(R{=}0,\tau{=}0)$ rises smoothly from 0.1 to 0.8, saturating near the pump laser coherence (Fig. 2d). This demonstrates that the system smoothly evolves into a coherent state above a critical pump power. The correlation function $N_{nm}$ directly maps to coherence $g^{(1)}(R{=}0,\tau{=}0)$ (SI Section S3) and is found to agree with our observations of $g^{(1)}(R{=}0,\tau{=}0)$ rising across the transition (theory, Figs. 2c,d). Because the Michelson probe integrates over multiple nanogaps within the diffraction limit, the measured $g^{(1)}(R{=}0,\tau{=}0)$ reflects spatially averaged mutual coherence.

The spatial expansion of $g^{(1)}(R,\tau{=}0)$ reflects the spread of correlations as dipoles are driven into inversion farther from the pump by the plasmon-enhanced transport between gaps. Similar behaviour is seen in thermalized photon Bose-Einstein condensates[39] and exciton-polariton condensates[40,41], where long-range order arises from relaxation into a quasi-equilibrium ground state with correlation lengths exceeding the system size. While details of spatial decay profiles vary depending on trap potentials, pump configurations, and system sizes, in common is a central highly coherent region, outside which is a near-exponential/power-law decay of quantum-degenerate low-energy states. Our strongly driven-dissipative system also shows a more complex spatial dependence (Fig. 2c), discussed further below.

The temporal decay of $g^{(1)}(R,\tau)$ shows an unusual coherence behaviour (Fig. 2e,f). Unlike the stretched exponential decay typical of exciton-polaritons[38] or exponential decay of photon BECs[42], $g^{(1)}(R,\tau)$ exhibits extended spatial coherence at $\tau$=0 (Fig. 2e) that rapidly decays within 10 fs. Additionally, the central coherence collapses and revives (Fig. 2f). This ultrafast dephasing arises from the high Purcell enhancement (>1000), so that dipoles, although spatially correlated, experience extremely rapid coherent decay.

The secondary rise in $g^{(1)}(R{=}0,\tau)$ is found to originate from spectral beating between multiple emission components, as confirmed by Fourier analysis of the measured spectrum (SI Fig. S21). However, phase-resolved interferometry reveals vortex nucleation near the first minimum of $g^{(1)}$, indicating that transient spatial phase reconfiguration also contributes to the observed revival. Thus, the collapse-revival-like feature reflects the interplay of spectral beating and spatial phase dynamics rather than coherent Rabi oscillations.

**Nonlinear Emission Dynamics and Universality**

The total emitted power initially scales linearly with pump power, then becomes sublinear at the onset of coherence (Fig. 3a). To account for variations of in-coupling efficiency across several hundred different pump positions, we rescale the in-coupled power at each position using its low-power linear regime as a reference (Fig. 3a, points; SI Section S5). Consistently across the sample, the averaged power dependence (open circles) reveals that the nonlinearity is influenced by $\Xi$, the number of molecules per gap.

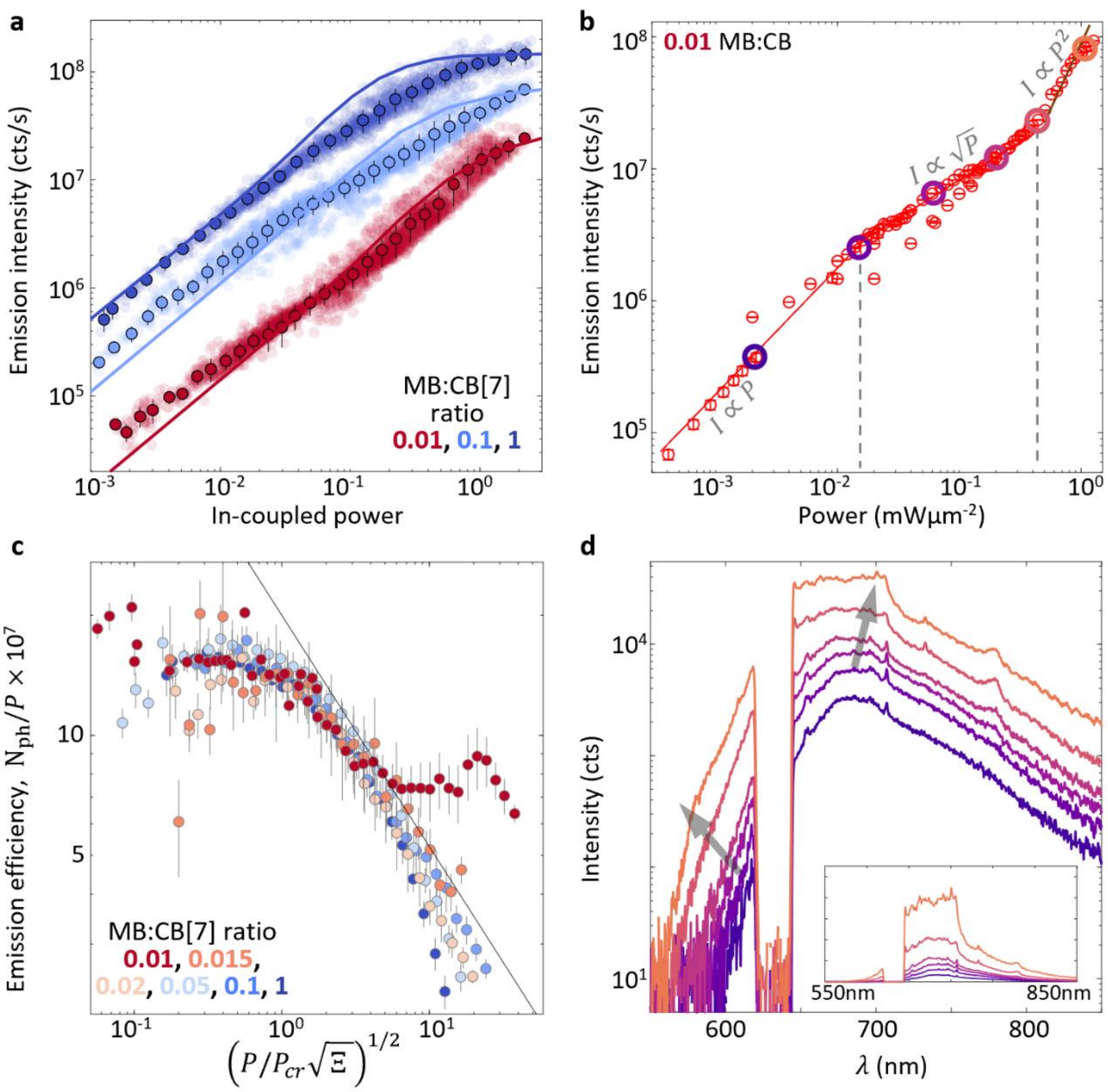

**Fig.3. Nonlinear emission for increasing emitter concentration.** (**a**) Integrated intensity of MB emission *vs* in-coupled power (details in SI Section S5) for increasing dye concentration, showing deviations from linear scaling for different regimes. Solid lines are theoretical predictions for the photon input-output curves. (**b**) Power dependence of emission intensity at one location showing transitions from linear to sublinear to super-linear regimes (solid lines indicate power laws as labelled). (**c**) Scaling of emitter efficiency ($I/P$) with increasing emitter density $\Xi$. Black line shows scaling as $(P/P_{cr}\sqrt{\Xi})^{-1/2}$ in intermediate regime above critical power $P_{cr} \propto \sqrt{\Xi}$. (**d**) Emission spectra from core region (corresponding to coloured circles in b). Inset shows spectra on linear scale. Arrows highlight non-linear growth of new Stokes/anti-Stokes emission components.

The sublinear core emission regime is consistently observed across all MB:CB ratios (Fig. 3a, solid arrows). While increasing from $\Xi \sim 2$ to 20 scales the emission by a factor of 10, further increasing to $\Xi \sim 200$ results in only a modest doubling of emission. For smaller $\Xi$, a superlinear emission region also emerges above a second threshold. The results of theoretically predicted input-output curves for photon number *vs* pumping rate match the experimentally observed data well (Fig. 3a), showing linear, nonlinear, and halo dominated- regimes arising from rapid plasmon-driven transport into the halo and saturation in the core (discussed below).

Detailed measurements of core emission *vs* power for low $\Xi$ at a single spot reveals three regimes (Fig. 3b): linear, $I \propto \sqrt{P}$ above a first threshold (dashed line), and $I \propto P^2$ beyond a second threshold $\sim$20 times larger. The exact value of power at which these thresholds occur varies from position to position (averaged results in Fig. 3a show smoother features, analysed further in SI Fig. S23). Emission is stable over time and the power dependence is consistently repeatable, indicating negligible bleaching since Purcell-accelerated radiative rates outpace intersystem crossing and oxidation. The scaling crossover reflects the evolving balance between collective coupling and nonlinear loss. Emission is linear at low pump, becomes sublinear as near-field coupling induces finite-range phase synchronization and halo formation, and at higher densities exciton-exciton annihilation competes with collective emission. Stronger annihilation at high MB:CB ratios suppresses superlinear growth, whereas reduced annihilation at low concentration ($\Xi$ = 0.01) enables it.

Tuning the dipole density within the nanogaps shows universal behaviour in the emission efficiency (photon intensity $I$ scaled to incoupled pump intensity $P$), with all input-output curves in Fig. 3a collapsing to the same curve in Fig. 3c, using input intensities scaled by the product of a critical intensity ($P_{cr}$) and the square root of the emitter density $\Xi$: as $(I/P) \propto (P/P_{cr}\sqrt{\Xi})^{-1/2}$. The exception to universality occurs where pair annihilation cannot happen, for the $\Xi$ = 0.01 case where there is on average only $\sim$1 emitter per nanogap.

Emission spectra (Fig. 3d) at increasing pump powers show no line narrowing, matching the temporally short $g^{(1)}(R=0,\tau)$, distinguishing this behaviour from conventional laser transitions. Low-power PL arises from the 0-0 and 0-1 vibronic transitions (modelled by Gaussians at 680 nm and 760 nm), along with sharp surface-enhanced Raman (SERS) peaks (SI Section S6). In the sublinear regime, where coherent spatial transport emerges, a new peak appears at 770 nm. At higher powers in the superlinear regime, additional Stokes and anti-Stokes emission emerges at 690 nm and 580 nm, resonant with the phonon energy difference between the 0-0 and 0-1 transitions. These vibrational features in the multi-peaked emission spectrum modulate $g^{(1)}$ via spectral beating (SI Fig. S21), reflecting the electronic-vibrational interplay intrinsic to molecular emitters and absent in atomic or conventional solid-state platforms. Such stimulated phonon driving has implications for the impact of optical pumping on phonon condensation/lasing[43], as well as optomechanical pumping of molecular vibrations[44].

**Coherence spreading and dipole-correlations**

Our model shows that the observed halo and coherence spreading arises from competition between light-matter coupling (generating coherence) and dipole-dipole pair annihilation (destroying it). At low pump strengths $\Gamma_n^{\uparrow}$, stimulated emission into plasmonic modes is insufficient to overcome losses ($\kappa$, $\Gamma^{\downarrow}$), so emission stays localized (Fig. 4a) and the dipoles in each gap remain uncorrelated.

As $\Gamma_n^{\uparrow}$ grows, emitter populations rise and greater pair-annihilation saturates the core, while stimulated emission transports coherence outwards via hopping $J$, forming a halo (Fig. 4b). The spatial pump profile creates a core with the highest excitation densities, where annihilation first suppresses emission. Coherence and emission are then redistributed outward via plasmonic coupling to regions of lower excitation density, forming an emission halo with enhanced width (Fig. 4b $N_{nn}$). Without plasmonic coupling ($t$=0), emission remains localized even at high pump strengths (Fig. 4b, grey). Removing pair annihilation ($\Gamma^{ee}$=0) leads to instability (SI Section S3), and the lack of any halo.

Although direct modulation of hopping $J$ via gap engineering is limited by the molecularly defined nanogap, systematic variation of Ξ allows controlled modification of $\Gamma_{\text{ee}}$, providing an experimental handle on the nonlinear term in the model. The observed concentration-dependent scaling in Fig. 3c therefore directly supports the theoretical assignment of pair annihilation as a key mechanism governing the collective regime.

Synchronization can be quantified by the rise in spatially-remote emitter correlations ($|Z_n|$) as $\Gamma_n^{\uparrow}$ increases (Fig. 4c), indicating phase locking and heralding the transition to the nonlinear regime where a halo appears (Fig. 4d). At the transition, on-site and off-diagonal coherences in the photon-photon correlations $N_{nm}$ and $g^{(1)}(R,\tau{=}0)$ grow in tandem (Fig. 2d and Fig. 4d inset), driven by dipole coupling. Here we thus observe a rapid crossover rather than a true phase transition. Our model shows that local emitter coherence $|Z_n|$ rises before photon correlations $N_{nm}$ spread spatially, establishing phase locking prior to halo formation. This sequential buildup explains why $g^{(1)}(R{=}0,\tau{=}0)$ increases at slightly lower pump intensities (Fig. 2d) than the full spatial expansion (Fig. 1h), reflecting a continuous evolution toward synchronicity. We note that extending this to a 2D model (Supplementary S3.7) gives comparable results.

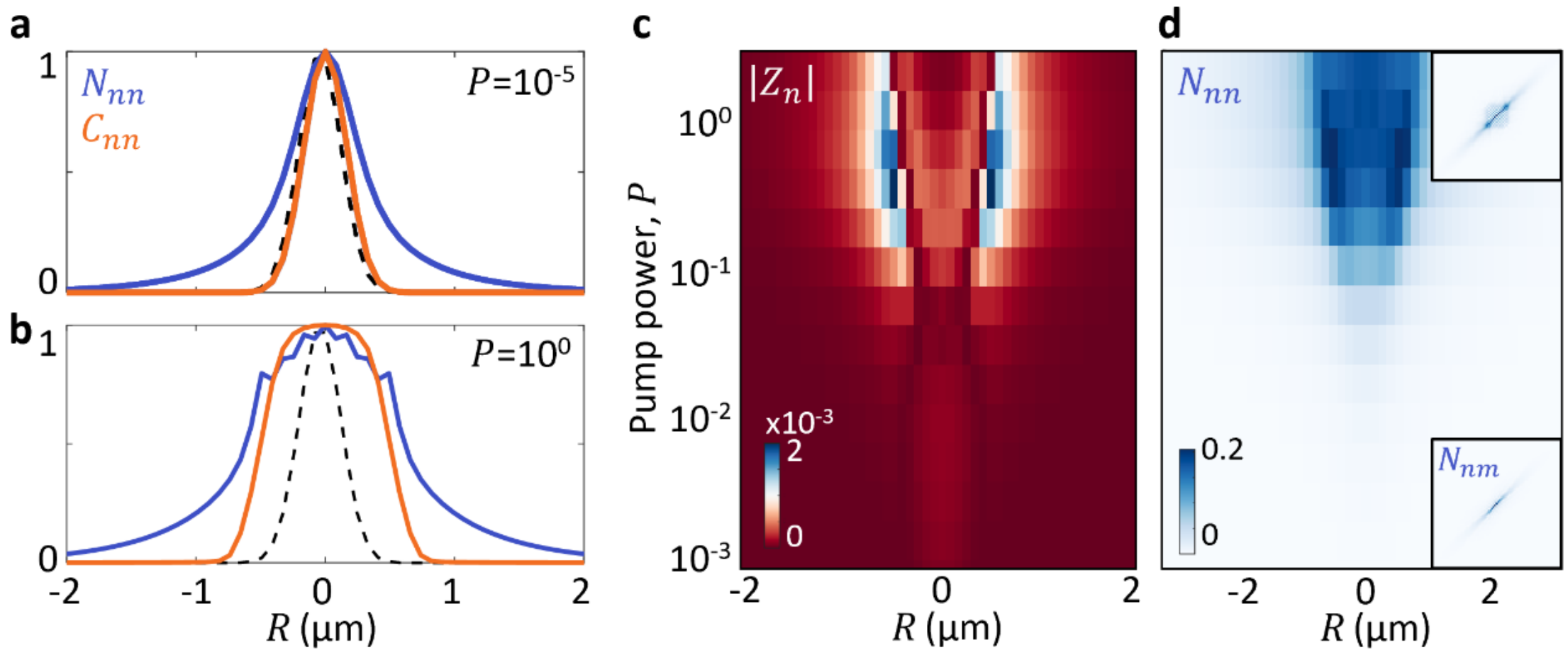

**Fig.4. Spreading of correlations and coherence.** (**a**,**b**) Theoretical model of normalized photon-emission profile ($N_{nn}$, blue) and on-site dipole correlations ($C_{nn}$, orange) for pump profile (black) at (a) low pump power $P$=$10^{-5}$ and (b) at high pump power $P$ =$10^{0}$, for $t$ = 0.4 eV and $\Gamma^{ee}$= $10^{-4}$ eV. (**c**) Model emitter-emitter correlations $|Z_n|$ and (**d**) photon emission ($N_{nn}$) as a function of pump power and position $R$. Inset shows the normalized off-site photon-photon coherences ($N_{nm}$) for pump powers in (a) (bottom inset) and (b) (top inset).

**Vortices and non-equilibrium behaviour**

Spatial coherence arises from dipoles synchronizing their phases through evanescent coupling (set by $J$ and $g$ in Eqn. (1)) and nonlinear dipole-nanocavity interactions (set by $\Gamma^{ee}$ in Eqn. (2)). However fast fluctuations from thermal/spontaneous emission noise and Purcell-enhanced decay, cause rapid temporal decoherence. The temporal decoherence manifests in the spatial domain as phase turbulence and slips, appearing as dynamically-evolving interference fringes, locally structured emission patterns, and vortices (Fig. 1f,5c). Figure 5b shows vortices appearing as fork-pair defects in the interferograms and as zeros in $g^{(1)}(x,y,\tau)$. Remarkably, Figure 5c shows that the presence or absence of vortices in the image changes with delay time.

The vortices represent points of zero visibility where the phase of the emission winds by 2π around a central defect. This is a topological signature that the system has established a degree of global phase order that is susceptible to topological excitations. The vortex here is a phase dislocation between successive frames of the optical field taken 11 fs apart (Fig. 5c). Note that since $g^{(1)}(x,y,\tau)$ involves time integration (at fixed delay $\tau$), this changing vorticity does not simply reflect a vorticity that changes with time; after time integration such fluctuating vorticity would simply give reduced coherence. Instead, this result indicates time-correlated fluctuations associated with changing vorticity.

These vortices only appear within the coherence time and resemble the time-averaged interferograms previously observed in polariton/photon condensates[45-47]. A vortex with $g^{(1)}$=0 traverses the central $R$=0, thus inducing the oscillations in $g^{(1)}$ seen in Fig. 2f and 5a. We suggest this puzzling behaviour of vortex symmetry breaking could be induced by disorder, in both spatial variation of plasmon transport and out-coupling rates, causing transient bursts of correlated emission to circulate as it propagates. Confirmation of the vortex phase winding around 2π, variation across sample positions, and agreement between experimental and simulated interferograms is discussed in SI Section S7.

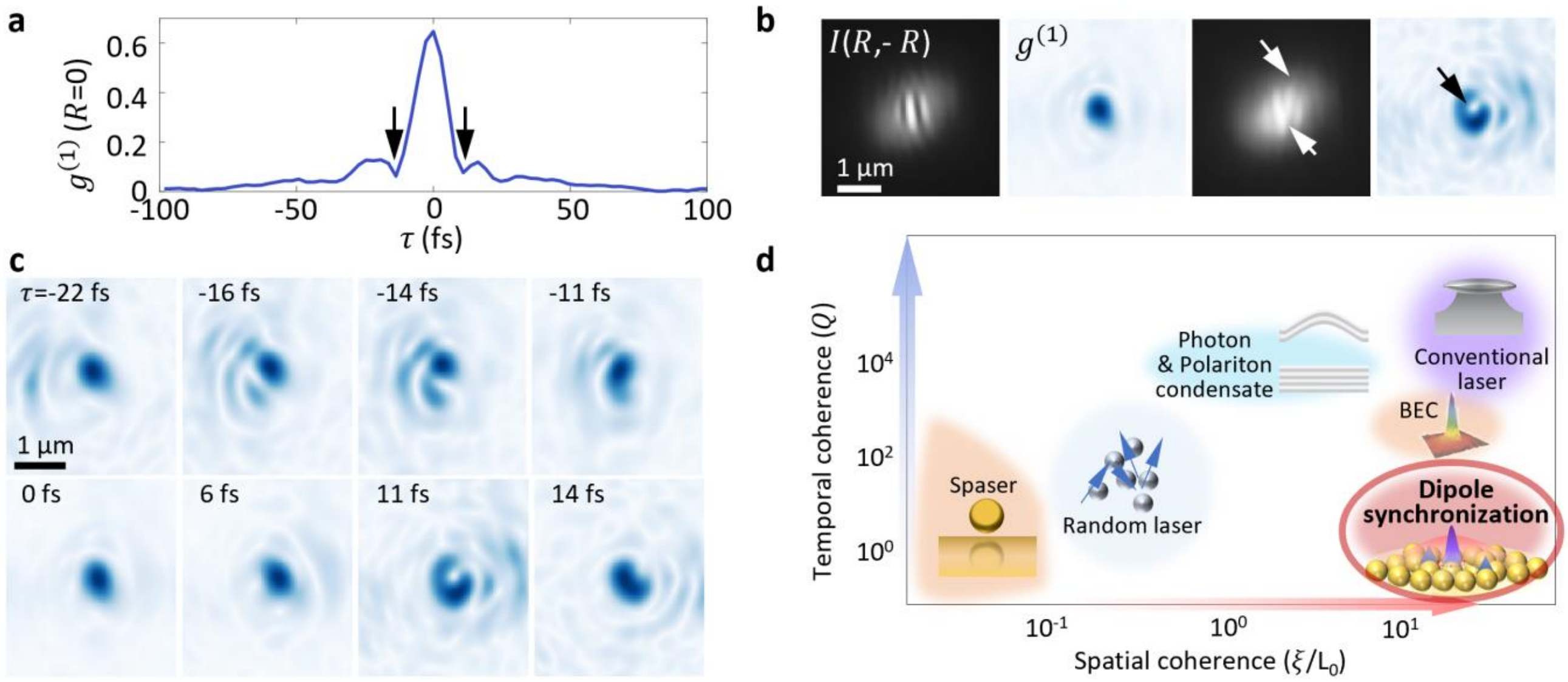

**Fig.5. Spatio-temporal vortices** forming as a function of delay time at pump power $P$=0.2 mW for MB:CB = 1 at different time delays $\tau$. (**a**) $g^{(1)}(R=0,\tau)$ coherence with minima corresponding to images below marked out. (**b**) Real-space Michelson interferogram intensity profile $I(R,-R)$ and extracted $g^{(1)}$ profiles for image with a vortex and without. Vortex indicated by an arrow as a zero in the $g^{(1)}$. (**c**) Vortex phase dislocation in the $g^{(1)}(x,y,\tau)$ in space and time, corresponding to zero observed in $g^{(1)}(R=0,\tau)$ coherence trace in (a). All images share the same scale bar. (**d**) Representative coherence properties of diverse synchronized systems (details in SI Section S10, SI Table 2). Temporal coherence denoted by quality factor of resonance or trap ($Q$) and spatial coherence by ratio of coherence length ξ to system size $L$.

## Discussion and Outlook

We have realized the first CW-pumped room-temperature synchronized dipole state. Emission, halo growth, and coherence onsets indicate phase locking among optical dipoles. The absence of spectral linewidth narrowing (Fig. 3d), together with rapid radiative and non-radiative decay, confirms short photon temporal coherence. Instead, coherence resides in molecular dipoles coupled via plasmonic near-fields. The large dipole-collective field coupling strength $g\sqrt{\Xi}$ > 100 meV arises from sub-nm gaps and extreme field confinement. Fast Purcell-enhanced emission limits temporal coherence, while lateral coherence reflects outward propagation of dipole phase correlations with phase diffusion setting the coherence length. The growth of dipole correlations manifests as an emission 'halo', which grows larger with a less focused pump (SI Fig. S40).

Synchronization in driven-dissipative dipole arrays has been theoretically established[7]. Our nanogap system contains the same ingredients - interacting two-level dipoles, collective decay, and incoherent pumping - within a strongly dissipative plasmonic architecture. Macroscopic coherence thus emerges from incoherent pumping as spatially structured emitter-emitter correlations. Our model reveals that local dipole spin-correlation ($Z_n$) establishes itself prior to the spatial expansion of photon emission (as the 'halo'). This sequential buildup distinguishes the synchronized dipole state from standard threshold-driven transitions in polaritons. Compared with other emissive systems (Fig. 5d, SI Table 2), synchronization is central: unlike random lasing, spatial coherence appears; unlike conventional lasing or polariton condensation[48], temporal coherence remains short; and unlike superradiance[49], operation is continuous-wave. In this extreme bad-cavity regime, phase order arises from near-field dipolar interactions on timescales shorter than dissipation, rather than from mode selection or equilibrium condensation.

Such driven synchronization without global temporal coherence is reminiscent of Josephson junction arrays[50], laser arrays[51], firefly swarms, neural networks, driven-dissipative ultracold atoms[52,53], and non-equilibrium 2D Bose condensates[54], exhibiting chimera-like coexistence of synchronized and asynchronous regions. Here, finite-range dipole phase locking extends over several coupled cavities and coexists with less correlated regions. The observation of spatio-temporal vortices indicates a unique regime of phase-turbulence and -slips that are inherently driven-dissipative and non-equilibrium, coexisting with localized synchronized regions.

In summary, the quasi-2D geometry of coupled nanogaps provides ultralow optical mode volumes, high Purcell factors, and strong interactions while enabling room-temperature CW operation. Disorder is intrinsic: dipoles are locally aligned with gap fields yet globally disordered, approximating a random gas. Synchronization manifests as spatial expansion and enhanced coherence rather than beamed emission or linewidth narrowing. We demonstrate that finite-range optical phase-locking emerges in this broadband, lossy, self-assembled nanophotonic environment, offering a scalable self-assembled route to new collective optical functions for sensing.

This platform advances mesoscopic quantum photonics at room temperature. The field of mesoscopic physics of few quantum emitters[6] is poised to push advances in physics and chemistry - we provide a new platform to realize this. Collective phase order among strongly coupled emitters enables engineered quantum resources [8,55], phase-coherent readout of embedded emitters such as molecular qubits or colour-centres, and routes to polariton-enhanced exciton transport and chemically active light-matter states [56,57]. More broadly, long photon lifetimes are not essential: in bad-cavity systems coherence can reside in emitters rather than the cavity field, improving metrology[58,59], while Purcell-enhanced ultrafast emission increases rates and can outpace slow environmental noise[60,61]. Synchronised dipolar dynamics, rather than high-$Q$ confinement, thus underpin collective optical behaviour in this nanoscale material.

## ACKNOWLEDGMENTS

We acknowledge funding from EPSRC (EP/L027151/1, EP/X037770/1 and EP/Y008162/1), and ERC (Project No. 883703 PICOFORCE and 861950 POSEIDON). R.A. acknowledges support from St. John's College Cambridge. P. F.-W. acknowledges support from EPSRC (EP/T518062/1). B. W. L. and J. K. acknowledge support from EPSRC (EP/T014032/1). The authors thank Prof. Riccardo Sapienza, Prof. Paivi Torma, Dr. T.V. Raziman, Prof. Rohit Chikkaraddy, Dr. Junyang Huang, and Dr. Lukas Jakob for discussions.

## AUTHOR INFORMATION

### Author contributions

R.A. and J.J.B. designed the experiments which were carried out by R.A. and L.B. Theory was developed by P.F.-W., B.W.L. and J.K. with input from R.A. and J.J.B. All authors contributed to the manuscript.

### Data availability

All the data in the current study are available from Cambridge Apollo repository, DOI assigned upon publication.

### Competing interests

The authors declare no competing interests.

### Additional information

**Supplementary information**. The online version contains available supplementary material.

Experimental Methods and Materials; Electromagnetic simulations of plasmonic chains; Interferometer calibration and coherence measurement protocol; Polariton vortex simulations and experiments; Power dependence and input power scaling; Surface-enhanced vibronic and Raman spectra; Theoretical model and calculations; Figs. S1 to S17; Table S1; References